\documentclass[modern,trackchanges]{aastex63}
\usepackage{todonotes}
\usepackage{epstopdf}

\usepackage{soul}
\usepackage{appendix}
\received{* * , 2020}
\revised{* * , 2020}
\accepted{* *, 2020}
\shorttitle{First SEP event seen on the Lunar Surface}
\shortauthors{Xu et al.}
\graphicspath{{./}{figures/}}

\begin{document}
\title{First Solar energetic particles measured on the Lunar far-side}

\correspondingauthor{Zigong Xu, Jingnan Guo}
\email{xu@physik.uni-kiel.de, jnguo@ustc.edu.cn}

\author[0000-0002-9246-996X]{Zigong Xu}
\affiliation{Institute of Experimental and Applied Physics, Kiel University, D-24118 Kiel, Germany}

\author[0000-0002-8707-076X]{Jingnan Guo}
\affiliation{CAS Key Laboratory of Geospace Environment, University of Science and Technology of China, Hefei 230026, China}
\affiliation{CAS Center for Excellence in Comparative Planetology, Hefei 230026, China}
\affiliation{Institute of Experimental and Applied Physics, Kiel University, D-24118 Kiel, Germany}
%
%
\author[0000-0002-7388-173X]{Robert \,F.\,Wimmer-Schweingruber}
\affiliation{Institute of Experimental and Applied Physics, Kiel University, D-24118 Kiel, Germany}
\affiliation{National Space Science Center, Chinese Academy of Sciences, Beijing, China}

\author[0000-0002-1390-4776]{Johan L.\,Freiherr von Forstner}
\affiliation{Institute of Experimental and Applied Physics, Kiel University, D-24118 Kiel, Germany}

\author[0000-0002-8887-3919]{Yuming Wang}
\affiliation{CAS Key Laboratory of Geospace Environment, University of Science and Technology of China, Hefei 230026, China}
\affiliation{CAS Center for Excellence in Comparative Planetology, Hefei 230026, China}

\author[0000-0003-3903-4649]{Nina Dresing}
\affiliation{Institute of Experimental and Applied Physics, Kiel University, D-24118 Kiel, Germany}

\author{Henning Lohf}
\affiliation{Institute of Experimental and Applied Physics, Kiel University, D-24118 Kiel, Germany}

\author{Shenyi Zhang}
\affiliation{National Space Science Center, Chinese Academy of Sciences, Beijing, China}
\affiliation{Beijing Key Laboratory of Space Environment Exploration, Beijing, China}
\affiliation{University of Chinese Academy of Science, Beijing, China}

\author{Bernd Heber}
\affiliation{Institute of Experimental and Applied Physics, Kiel University, D-24118 Kiel, Germany}

\author{Mei Yang}
\affiliation{Beijing Institute of Spacecraft System Engineering, Beijing, China}
%
%



\begin{abstract}
On May 6, 2019 the Lunar Lander Neutron \& Dosimetry (LND) Experiment on board the Chang'E-4 lander on the far-side of the Moon detected its first solar energetic particle (SEP) event with proton energies up to 21 MeV. Combined proton energy spectra are studied based on the LND, \emph{SOHO}/EPHIN and \emph{ACE}/EPAM measurements which show that LND could provide a complementary dataset from a special location on the Moon, contributing to our existing observations and understanding of space environment. We applied velocity dispersion analysis (VDA) to the impulsive electron event and weak proton enhancement and show that electrons are released only 22 minutes after the flare onset and $\sim$15 minutes after the type II radio burst, while protons are released more than one hour after the electron release. 
The beam-like in-situ electrons and clear velocity dispersion indicate a good magnetic connection between the source and Earth. This is remarkable because stereoscopic remote-sensing observations from Earth and STEREO-A suggest that the Solar energetic particles (SEPs) are associated with an active region nearly 113$^\circ$ away from the magnetic footpoint of Earth.  
This suggests that these SEPs did not propagate along the nominal Parker spiral normally assumed for ballistic mapping and that the release and propagation mechanism of electrons and protons are likely to differ significantly for this event. 

\end{abstract}

\keywords{Solar energetic particles, Space Radiation, Lunar Exploration}


\section{Introduction} \label{sec:intro}

The Chang'E-4 mission, which consists of a lander, a rover and a relay satellite, is the first mission to land on the far side of the Moon. It landed in the von\,K\'arm\'an Crater on January 3, 2019, 02:26 UTC. The Lunar Lander Neutron and Dosimetry experiment (LND) \citep{Wimmer-Schweingruber2020} on board the lander of Chang'E-4 is designed to take active dosimetry measurements on the surface of the Moon as its chief scientific goal. Apart from the primary objective of LND which is to measure the radiation level on the lunar far-side preparing for astronaut missions \citep[][]{zhang-etal-2020}, the charged particle telescope also provides high-quality data of energetic particles and contributes to heliophysics. For example, LND provides proton and Helium-4 spectra between 9 MeV/nuc and 35 MeV/nuc.

During the first year of the mission (2019), solar activity was minimal and LND detected only two small SEP events on May 4 and 6, 2019, of which the second had sufficient counting statistics for protons between 9.0 and 21.0 MeV to allow a meaningful analysis. This event was related to an active region located at E50, nearly 113$^\circ$ away from the Earth's nominal coronal magnetic footpoint, where an M1.0 class flare erupted before the SEPs onset and a narrow and slow coronal mass ejection (CME) appeared later.
Combining remote sensing observations of the solar source with in-situ particle measurements from multiple spacecraft, we analyze and discuss the possible particle release and transport processes. 

\section{Observations}  \label{sec:Observation}

\subsection{In-situ measurements} \label{sec:In-situ measurementm}
\begin{figure}
    \centering
    \includegraphics[width=0.9\textwidth,height=0.6\textwidth]{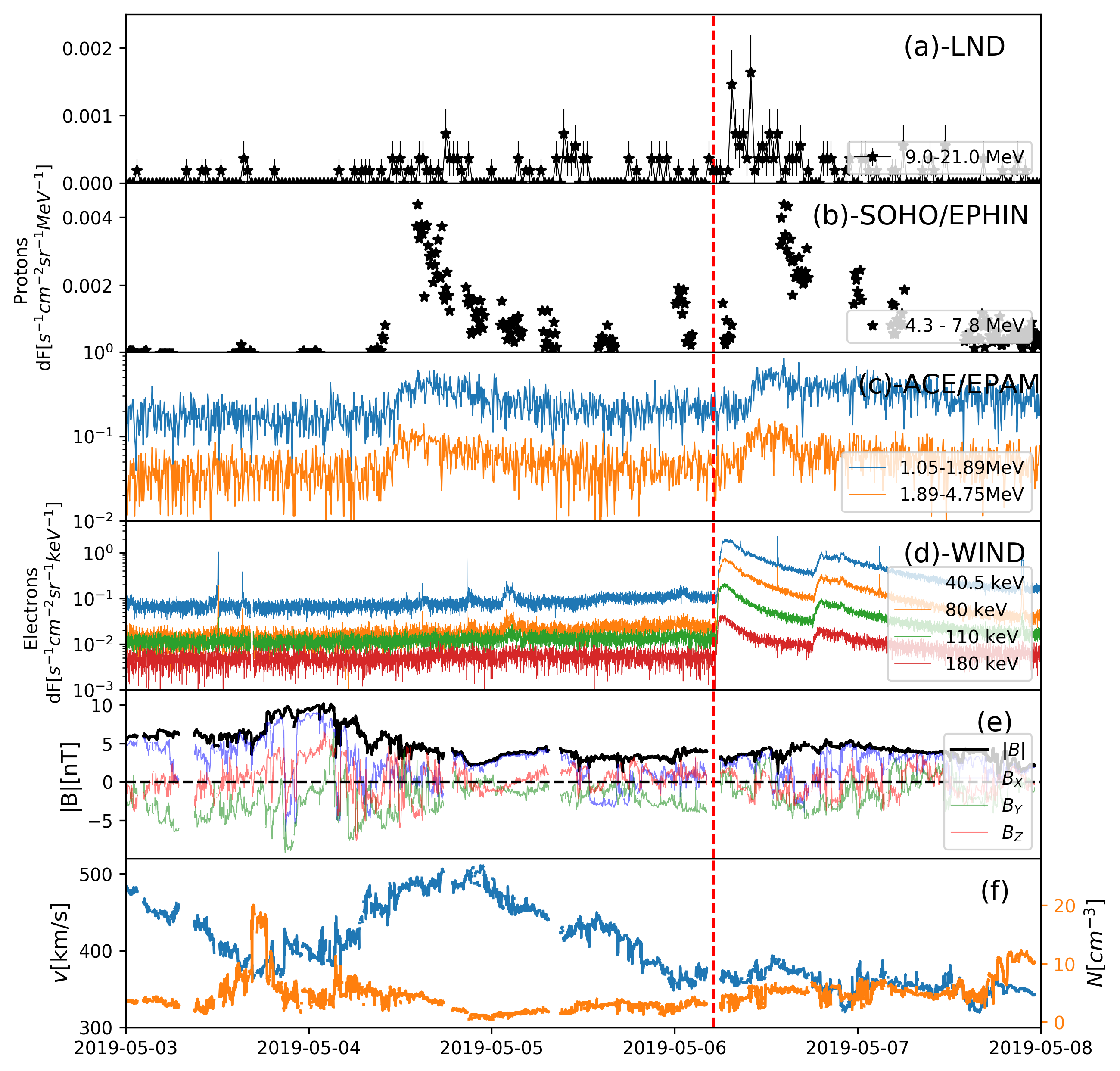}
    \includegraphics[width=0.3\textwidth,height=0.3\textwidth]{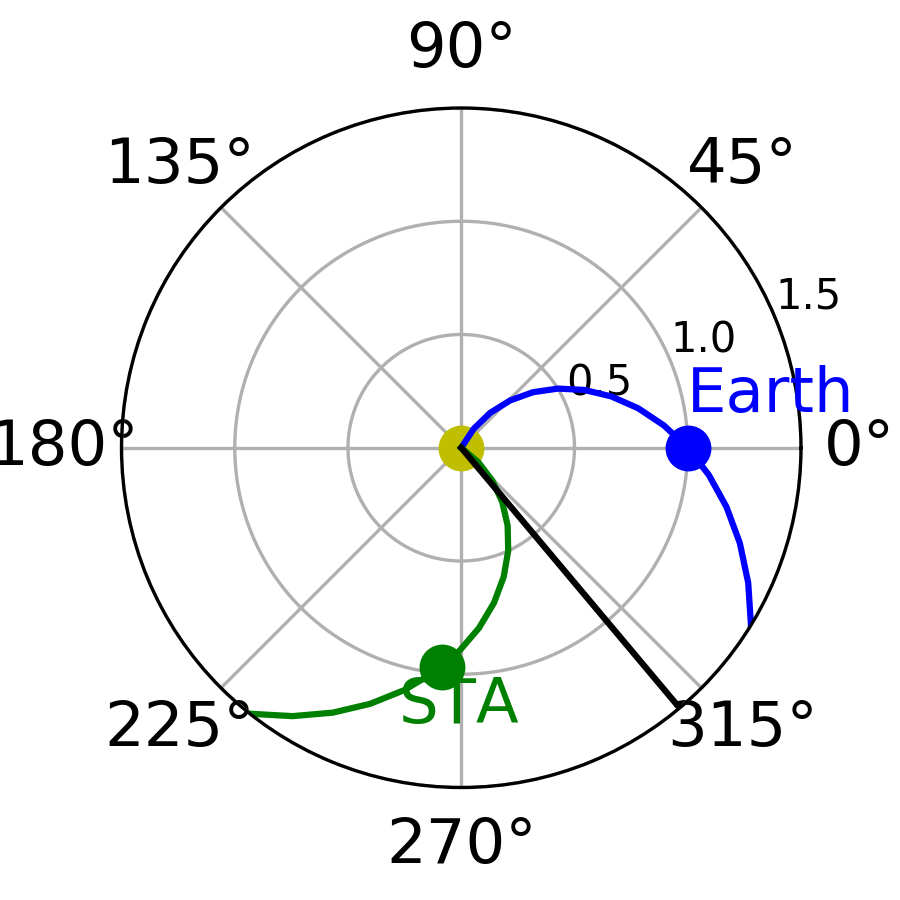}
    \includegraphics[width = 0.6\textwidth,height =0.3\textwidth]{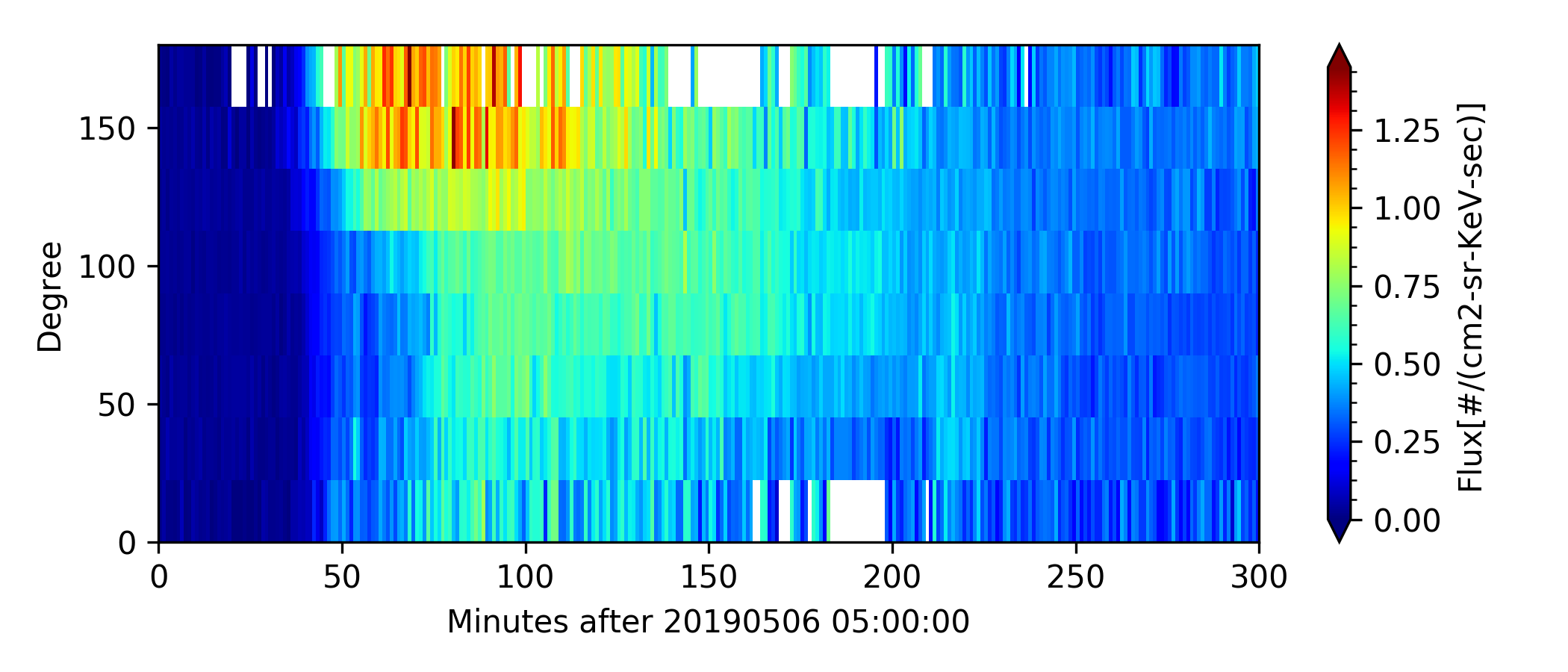}
    \caption{In-situ measurements near Earth, including (a) proton intensity at 9.0-21.0 MeV by \emph{LND}, (b) proton intensity at 4.3-7.8 MeV by \emph{SOHO/EPHIN}, (c) ion intensity in 1.05-4.75 MeV by LEMS120 of \emph{ACE/EPAM}, (d) electron intensity in 40.5 keV-180 keV by \emph{Wind/3DP}, (e) magnetic field magnitude and three components in the GSE coordinates, (f) solar wind velocity and plasma density from OMNI.
    The red line indicates the eruption time of solar flare on May 6. The radial direction of the active region as well as the nominal Parker spirals for Earth and STA are plotted in the HEEQ coordinate in the bottom left panel. The pitch angle distribution of electrons measured by \emph{Wind} after 05:00:00 on May 6 is given in the bottom right panel. Electrons are flowing outwards towards the observer at 180 degrees.}
    \label{fig:L1_measurement}
\end{figure}
Chang'E 4 landed on the far side of the Moon and can only operate during its local daytime because it is too cold at night. Therefore, LND only provides measurements ahead of the Earth's bow shock. It measures the energy which a particle deposits in its 10 detectors. LND can stop and thus identify the species and energy of charged particles up to 30 MeV/nuc. In order to better understand the temporal variation of protons and electrons at other energies during the event, we also include observations from the Electron Proton Helium Instrument \citep[EPHIN,][]{muller1995costep} on board \emph{SOHO}, the Electron Proton Alpha Monitor \citep[EPAM,][]{Gold1998SSRv} on board the Advanced Composition Explorer (\emph{ACE}), as well as the 3-D Plasma and Energetic Particle Investigation \citep[3DP,][]{lin1995three} on board \emph{Wind}. Those data are presented in the top panel of Fig.~\ref{fig:L1_measurement} and are discussed from top to bottom in the following paragraph.

After the flare eruption at 04:56 on May 6 (Tab.~\ref{tab:Time-line}, more details in Section \ref{sec:Remote-sensing}) which is marked as a red vertical dashed line in Fig.~\ref{fig:L1_measurement}, LND detected the arrival of SEPs as shown in panel (a) of Fig.~\ref{fig:L1_measurement}.
The proton channel 9.0-21.0 MeV shown here is the combination of the first five channels of LND's one-minute proton data which are provided in appendix. Due to the low intensity of the event and poor counting statistics in the data, the flux is averaged over 30 minutes for this figure. The analysis described below will be performed using the highest possible time resolution. In panel (b) we display the 4.3--7.8 MeV proton intensity profile measured by \emph{SOHO}/EPHIN. However, the EPHIN measurements have data gaps due to limited telemetry during a \emph{SOHO} roll maneuver on May 6. Thus, LND becomes the only instrument that observed the complete duration of this event at these energies.

Both y-axes in panels (a) and (b) are plotted in linear scale in order to better show the SEP enhancement. Panel (c) shows the ion flux averaged over each 30 minutes for two energy channels between 1 and 4.75 MeV detected by the LEMS30 (Low-Energy Magnetic Spectrometer) detector, one of the telescopes of \emph{ACE}/EPAM. 
The intensity profile shows clear velocity dispersion of ions which are mostly attributed to protons.


Finally the electron profile observed by \emph{Wind} between 40 keV and 180 keV is plotted in panel (d) which clearly shows electrons starting to arrive at around 5:30 on May 6, slightly earlier than the energetic protons. Different energy channels show different onset times with higher energy electrons arriving earlier. This velocity dispersion feature is studied using a velocity dispersion analysis (VDA) method as discussed below.
The second increase later that day is associated with another solar eruption which is not considered in this study.

The remaining panels of Fig.~\ref{fig:L1_measurement} show the local solar wind plasma data from 1-minute Near-Earth Heliosphere Data(OMNI) including the magnetic field, solar wind speed, and proton density. The near-Earth space is rather calm around the time of the SEP event of interest, with no indication of transient Interplanetary Coronal Mass Ejections (ICMEs), shocks or stream interaction regions (SIR) which could otherwise contribute to possible local acceleration of particles around the onset of the SEP event.A preceding SIR on May 3/4 is indicating by the increase of magnetic fields, enhanced proton density and the slow rise of solar wind speed and was followed by a high-speed stream on May 4/5. We don't expect this preceding structure to have any influence on the SEP event reported here.

As shown in the bottom left panel of Fig.~\ref{fig:L1_measurement}, at the time of the SEP event under study, the longitudinal separation between the active region and the magnetic footpoint of STA using ballistic back mapping is only 7$^\circ$ which suggests that STA would be a perfect observer for these energetic particles. However, STA was already experiencing an ongoing SEP event at the time of the flare and no impulsive contribution is seen at STA. Thus all we can state is that the impulsive event reported here was too small to be seen above the background of the preceding event at STA.

\begin{figure}
    \centering
    \includegraphics[width=\textwidth,height= 0.5\linewidth]{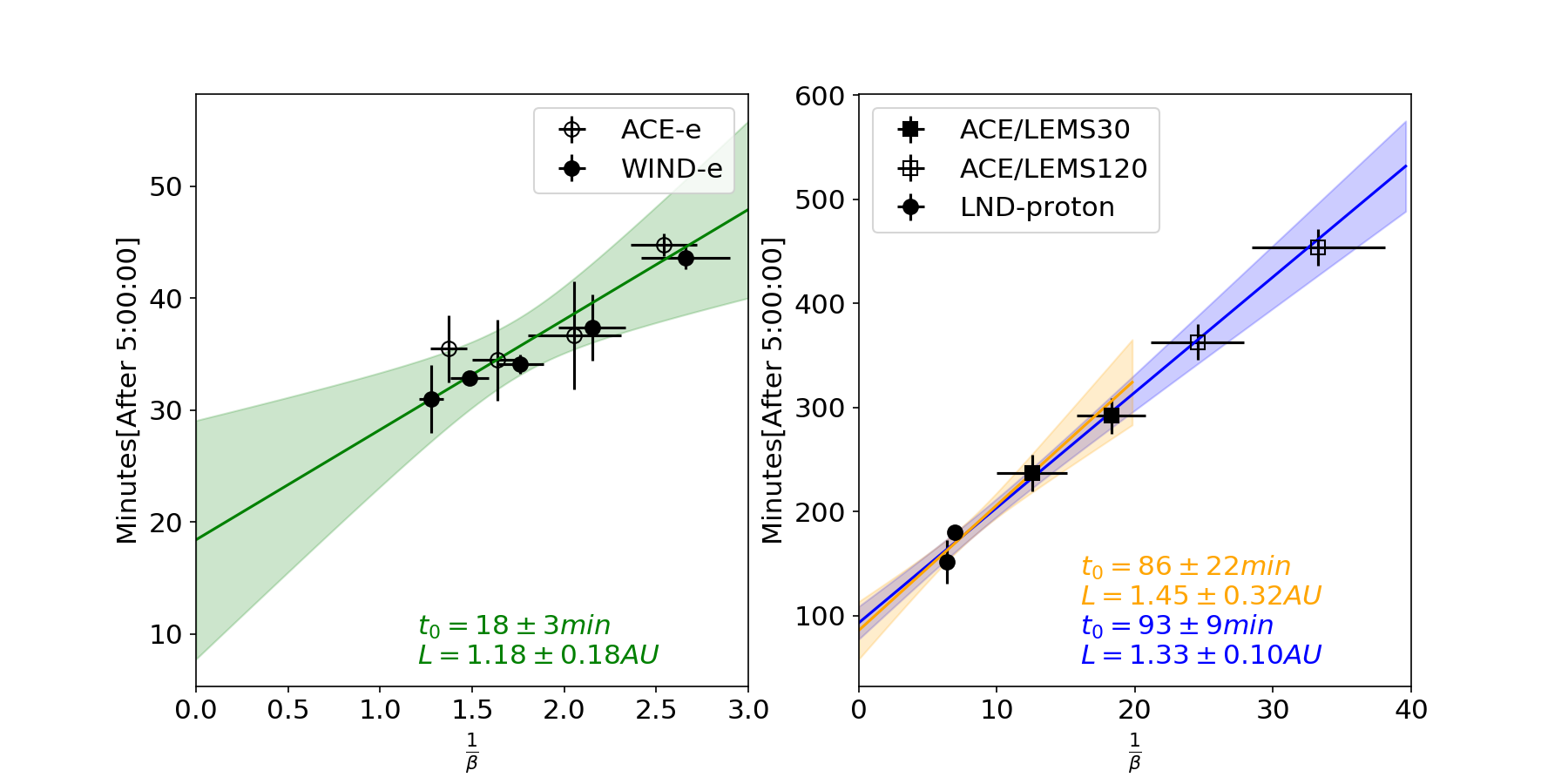}
    \caption{Velocity dispersion analysis of the SEPs (electrons shown in the left and protons shown on the right panel) on May 6, 2019. \emph{Wind} and \emph{ACE} electron data are used to determine the electron release time. LND proton, \emph{ACE}/EPAM LEMS30 and LEMS120 ion data are used to determine the proton release time. The linear fits and the fitted parameters of the VDA analysis with 95\% confidence interval are marked in the plots. More details can be found in the text of Sec.~\ref{sec:VDA}. }
    \label{fig:VDA}
\end{figure}

\subsection{Determination of onset and release times}\label{sec:VDA}
The small intensity of the SEP event along with the limited geometry factor of LND require careful analysis of the data and limit the accuracy of the determination of the event onset times.
Therefore, we apply the Poisson-CUSUM method \citep{lucas1985-CUSUM} and follow the procedure in \citep{Huttunen-Heikinmaa-2005A&A} to derive the onset time of each energy channel using LND's highest time resolution of 1 minute. In the appendix, we present the Poisson-CUSUM analysis on the different LND energy channels in more detail. 

The onset times of the 9.0--10.6 MeV and 10.7--12.7 MeV channels are 08:00 $\pm$ 7 min and 07:32 $\pm$ 21 min. The calculation of the uncertainties is explained in the appendix. Assuming protons travelling scatter-free along a 1.2 AU interplanetary magnetic field (IMF) line, which is calculated from the solar wind speed averaged over 8 hours before the SEP event, we derive that the 10.7--12.7 MeV protons need about 63 minutes to arrive at Earth and the release time is around 06:29 $\pm$ 21 min. The onset times for \emph{ACE}/EPAM and \emph{Wind}/3DP are determined as the time when the flux exceeds $3\sigma$ of the background signal. The latter is defined as the flux an hour before the SEP onset. 

In Fig.~\ref{fig:VDA}, the onset times $t_{onset}$ of different channels are plotted versus $1/\beta$ that is $c/v$, where $c$ is the speed of light and $v$ is the speed of particles with different energies. The uncertainties of the onset times at \emph{ACE} and \emph{Wind} are computed as the difference between the onset times using a 1-$\sigma$ threshold and a 3-$\sigma$ threshold. The error bar in the $1/\beta$ direction is calculated from the width of the energy channels. Both \emph{ACE}/EPAM and \emph{Wind}/3DP onset times of electrons are plotted in the left panel of Fig.~\ref{fig:VDA}.  Proton onset times are plotted in the right panel with the \emph{ACE/EPAM} LEMS30 channels shown as filled squares, LEMS120 as empty squares and high-energy LND channels are shown as filled circles. Because only two channels of LEMS30 are available, two additional channels from LEMS120 with lower energy are also used here.

We apply the VDA method using the measured onset times in different energy channels to determine the particle release time, $t_0$ and the length of the IMF spiral, $L$, along which particles propagated. We fit the measured onset times $t_{\rm onset}$ and inverse velocities, $1/v$, with the function $t_{onset} = t_{0} + L/v$ using orthogonal distance regression (ODR) to account for uncertainties in both the $x$ and $y$ directions. The underlying assumption is that the particles which arrive earliest have undergone little scattering. The linear VDA fits are plotted as solid lines in the left and right panels of Fig.~\ref{fig:VDA} for electrons and protons, respectively, and 95\% confidence intervals are given as shaded regions. The fitting result based on protons from the LND and LEMS30 measurements is given in orange, while the result using all data is given in blue. The two fit results are consistent. The results for $t_{0}$ and $L$ are also given in Tab.~\ref{tab:Time-line}.

The results of the fits suggest that electrons are released $22 \pm 3$ minutes after the flare eruption and that protons are released about $75\pm 12$ minutes after the electron release. The IMF length inferred from electrons is $L_e = 1.18 \pm 0.18$ AU which is consistent with the Parker spiral derived from the in-situ solar wind speed, $L = 1.2$ AU. For protons we obtain a length $ L_p = 1.34\pm 0.10$ AU. $L_e$ and $L_p$ are consistent with each other and weighting them with their inverse errors we find an average $L = 1.28 \pm 0.15$ AU.


The VDA method assumes that all particles stream along the same IMF which may become invalid in a realistic condition. In order to assess the reliability of this assumption for this SEP event, we also checked the pitch-angle measurement by \emph{Wind}. At the beginning of the SEP event, electrons show a clear anisotropy (enhanced intensity at $\sim180^\circ$ pitch angle, bottom right panels of Fig.~\ref{fig:L1_measurement}) suggesting that electrons first arrived along the IMF (Tab.~\ref{tab:Time-line}). 
Consequently, these particles experienced little scattering, in agreement with the requirement of the VDA method. 
Unfortunately, proton observations with much smaller statistics make it difficult to identify a significant anisotropy of protons from \emph{Wind}. 




\subsection{Energy Spectra}\label{sec:spectra}
\begin{figure}
    \centering
    \includegraphics[width=\textwidth]{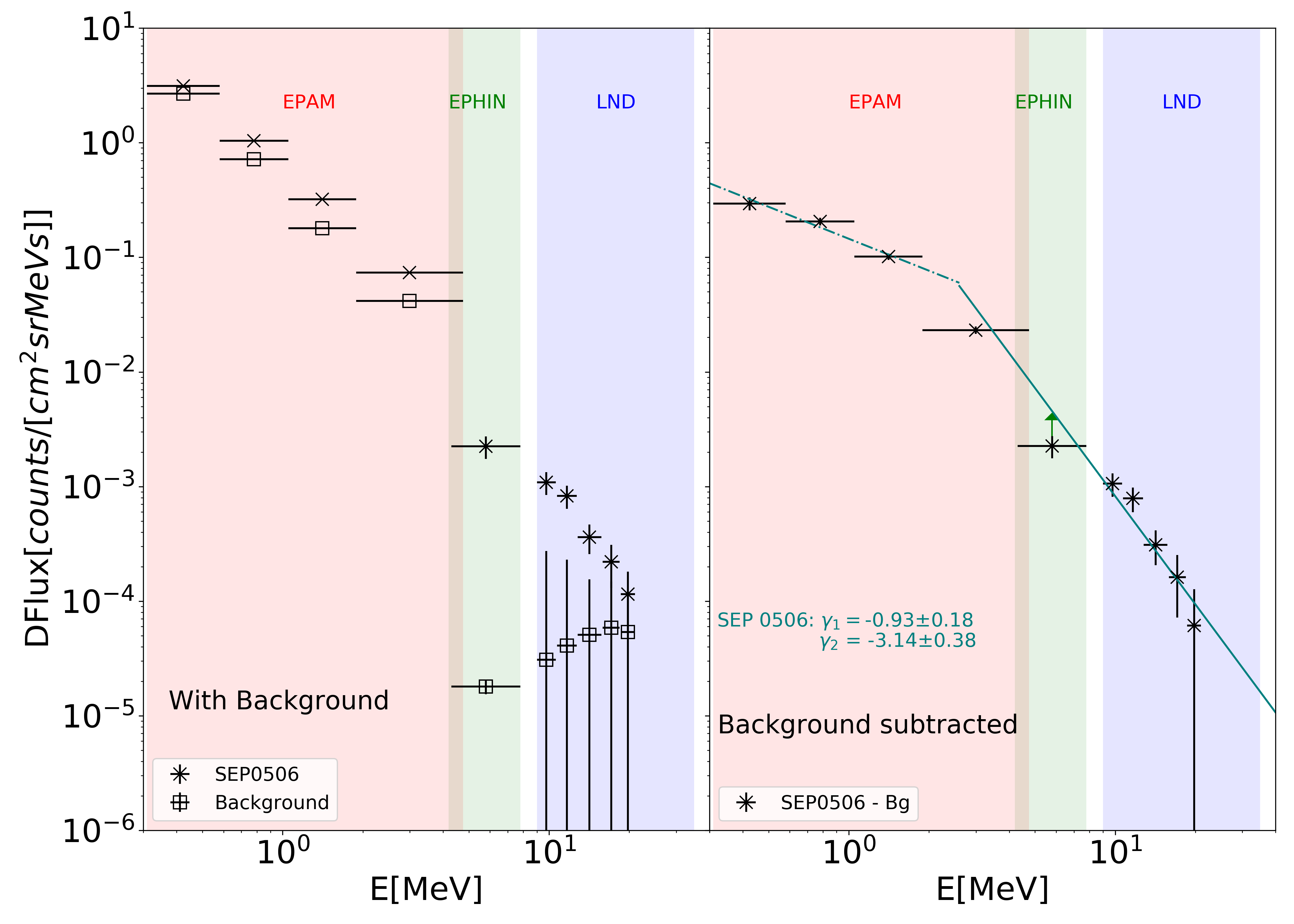}
    \caption{Proton spectra of the SEPs on May 6, 2019 and the background spectra which are averaged between April 29 and May 3. The left panel shows the spectra including the background and the right panel shows the SEP spectra with background subtracted. Different shaded areas indicate the energy ranges covered by the different instruments from $\sim$300 keV to $\sim$30 MeV. More details can be found in the text of Sec.~\ref{sec:spectra}.}
    \label{fig:LND_SEP_Spectra}
\end{figure}

Combining the proton measurement by \emph{SOHO}/EPHIN as shown in panel (b) of Fig.~\ref{fig:L1_measurement} and the ions registered by \emph{ACE}/EPAM in panel (c) which are dominated by lower energy protons and also those of LND, we plot the proton spectra integrated between 06:00 and 17:00 on May 6, 2019 in the left panel of Fig.~\ref{fig:LND_SEP_Spectra}. 
The areas shaded in blue, green, and pink are the energy coverage of LND (9 MeV--35 MeV), \emph{SOHO}/EPHIN (4.3 MeV--7.8 MeV) and \emph{ACE}/EPAM (0.310 MeV--4.75 MeV) respectively. 
The background spectra are also plotted and they are averaged between April 29 and May 3 (LND started working on April 29 for its fifth lunar day measurement). 
The energetic proton spectra during the event are well above the background as shown in the left panel of Fig.~\ref{fig:LND_SEP_Spectra}, despite of the large uncertainty in the LND data due to the low number of counts.
The right panel of the figure shows the SEP spectra where the background spectrum has been subtracted. 
Because of the EPHIN data gaps during the impulsive phase of the SEP event as shown in Fig.~\ref{fig:L1_measurement}, the flux of the SEP event measured by EPHIN is likely to be larger than the derived one, as pointed out by the upward-pointing green arrow in the right panel. 


Despite this event being of a more impulsive nature, we use a classic double power-law spectrum to fit the data. As can be seen in Fig.~\ref{fig:LND_SEP_Spectra}, there is a break in the spectral slopes around 2.5 MeV, a fit using a double power law gives power-law indices of $-0.93\pm0.18$ and $-3.14\pm0.38$ for the spectra below and above this break energy, respectively. Interestingly, the break energy falls right between the values for large gradual events \citep{mewaldt2012energy} and those determined for very small events by \cite{Joyce2020ApJS} using data from Parker Solar Probe \citep{Fox2016SSRv}. 

\subsection{Remote-sensing observations} \label{sec:Remote-sensing}
\begin{deluxetable}{l c c r}
\tabletypesize{\scriptsize}
\tablecaption{Time line of the SEP event on 6 May 2019}
\tablehead{
\colhead{Time(UT)}&
\colhead{Event}&
\colhead{Characteristics}&
\colhead{Refer to}\\
\colhead{2019.05.06}
}
\startdata
04:56:00 &Eruption of Flare and start of SXR& AR 12740@50$^\circ E;$ M1.0 & \\
& & $D_{E}$ =$113^\circ; D_{STA} = 7^\circ$  & \\
04:57:23 & Type III radio burst & SSRT, \emph{Wind}/WAVES, \emph{STA}/SWAVES &   \\
05:00:22 & EUV wave propagating toward west & AIA 193\AA; $V\approx500km/s$ & Fig. \ref{fig:connection} \\
05:03:30 & Type II radio burst & $@230MHz-90MHz$, Last for 6 minutes &  \\
05:18:24 & Electron release & VDA, IMF$\sim$1.18AU, beam-like & Fig. \ref{fig:VDA} \\
05:28:00 &CME first appears at SOHO/LASCO C2 & 376km/s\tablenotemark{a}; 326km/s\tablenotemark{b}& Fig. \ref{fig:connection}\\
05:30:59 & 310keV electron onset & \emph{Wind} & Fig. \ref{fig:VDA} \\
05:31:00 & CME at STA COR2& AW$\sim20^\circ$, deflect to west& Fig. \ref{fig:connection}\\
06:27:00 & 10.7-12.7MeV proton release & LND; TSA & Fig. \ref{fig:VDA}\\
06:29:42 & Proton release & VDA, IMF$\sim$1.33AU & Fig. \ref{fig:VDA} \\
07:32:00 & 10.7-12.7MeV proton onset &LND, Poisson-CUSUM \citep{lucas1985-CUSUM} &Appendix
\enddata
\tablecomments{The 8.3 minutes light travel time has been subtracted from all the times in this table.\\
$D_E$ is the longitudinal distance between active region and magnetic footpoint of the Earth; $D_{STA}$ is the distance to the STA footpoint; SSRT = Siberian Solar Radio Telescope; AW = angular width; GOES = Geostationary Operational Environmental Satellite; VDA = velocity dispersion analysis; IMF = interplanetary magnetic field; TSA = Time shifted analysis}
\tablenotetext{a}{Velocity from Cactus catalog}
\tablenotetext{b}{Velocity from GCS fitting}
\label{tab:Time-line}
\end{deluxetable}

On 2019 May 6, an impulsive M1.0 flare erupted from active region (AR) 12470 located at N08E50. The soft X-ray (SXR) flare had an onset at 04:56:00\footnote{The time for remote-sensing measurements in this study has subtracted the $\sim8.3$ minutes light travel time over 1 AU distance.} and lasted for $\sim$8 minutes as detected by the solar X-ray Imager (SXI) on the GOES satellite.
Since this is the only visible eruptive source on the Sun seen from Earth before the onset of the SEPs, we believe this is the solar counterpart of the SEPs measured in-situ near Earth and Moon.  
Nearly at the same time of the SXR emission, a broadband type III radio burst starting from 240 MHz was observed by not only the ground radio observatory Solar Radio Telescope (SSRT\footnote{\url{http://www.e-callisto.org/}}) on Earth but also by the WAVES instrument on board the \emph{Wind} spacecraft at Earth L1 point as well as by WAVES on board STEREO-A (STA). A type II radio burst between 230--90 MHz indicating the existence of a coronal shock was reported by the SSRT during 05:03:30--05:09:30. 
The timeline of this event is given in Tab.~\ref{tab:Time-line}.

\begin{figure}[!ht]
    \centering
    \includegraphics[width = \textwidth]{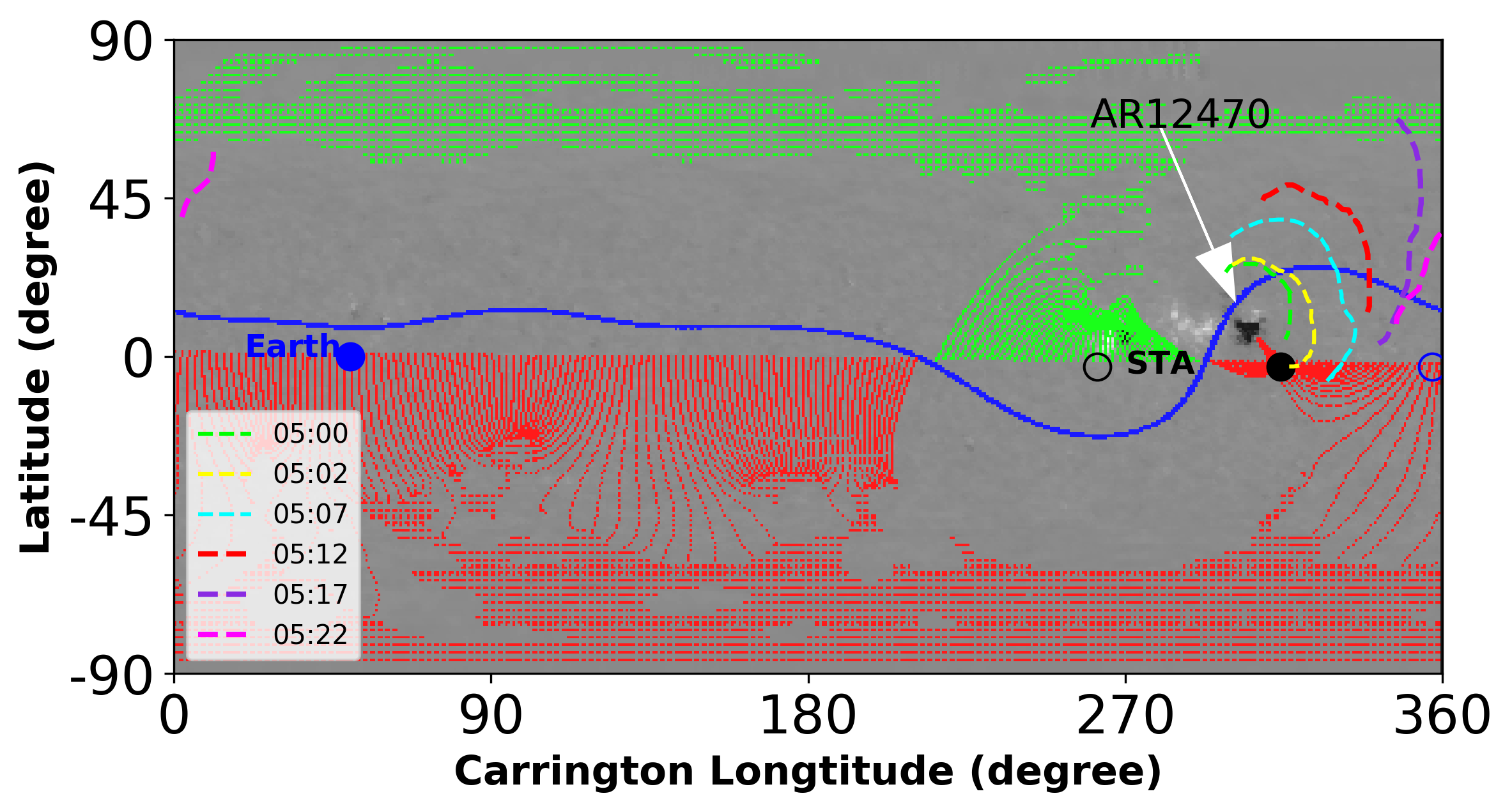}
    \includegraphics[width = 0.46\textwidth]{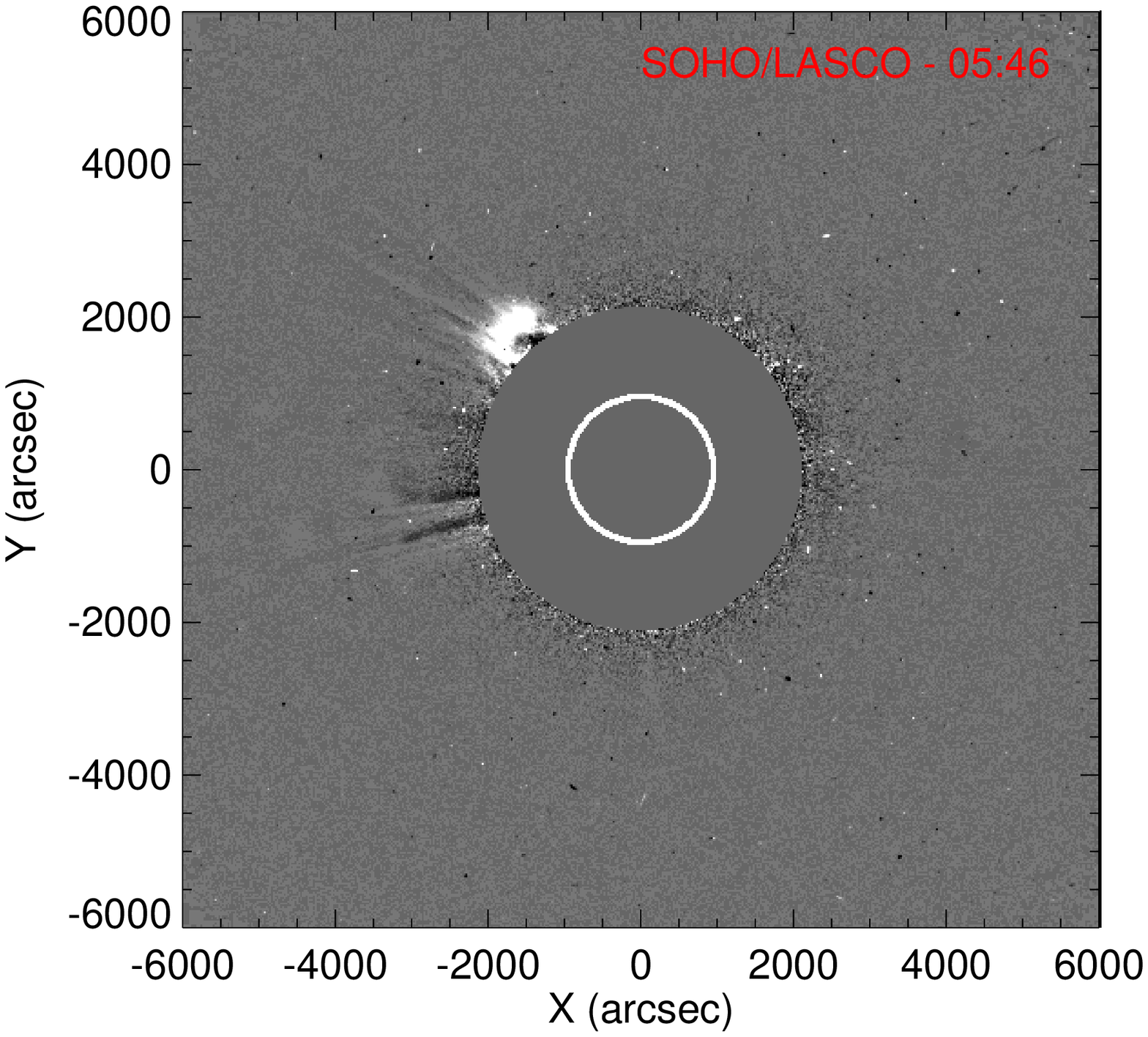}
    \includegraphics[width = 0.46\textwidth]{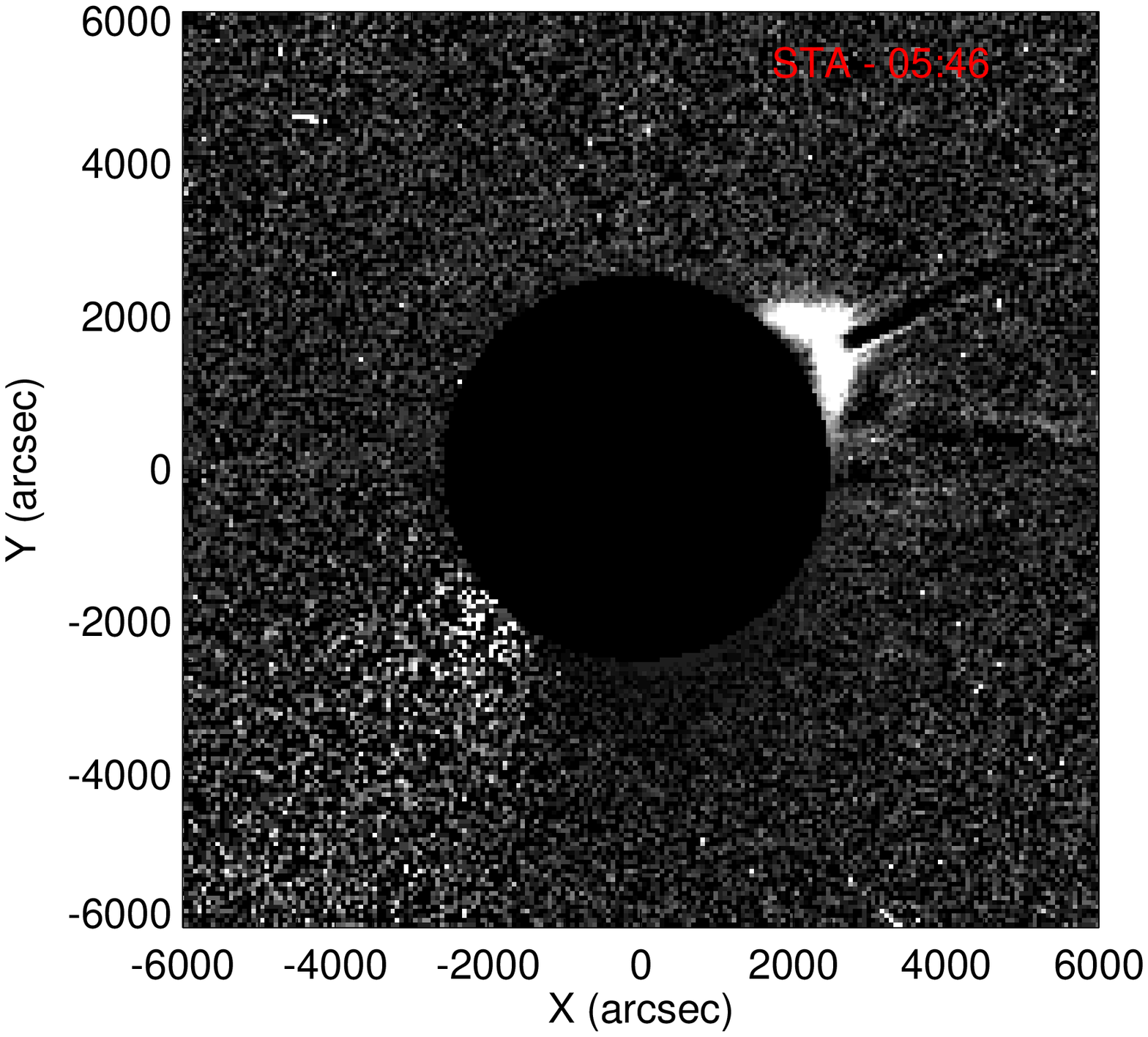}
    \includegraphics[width = 0.46\textwidth]{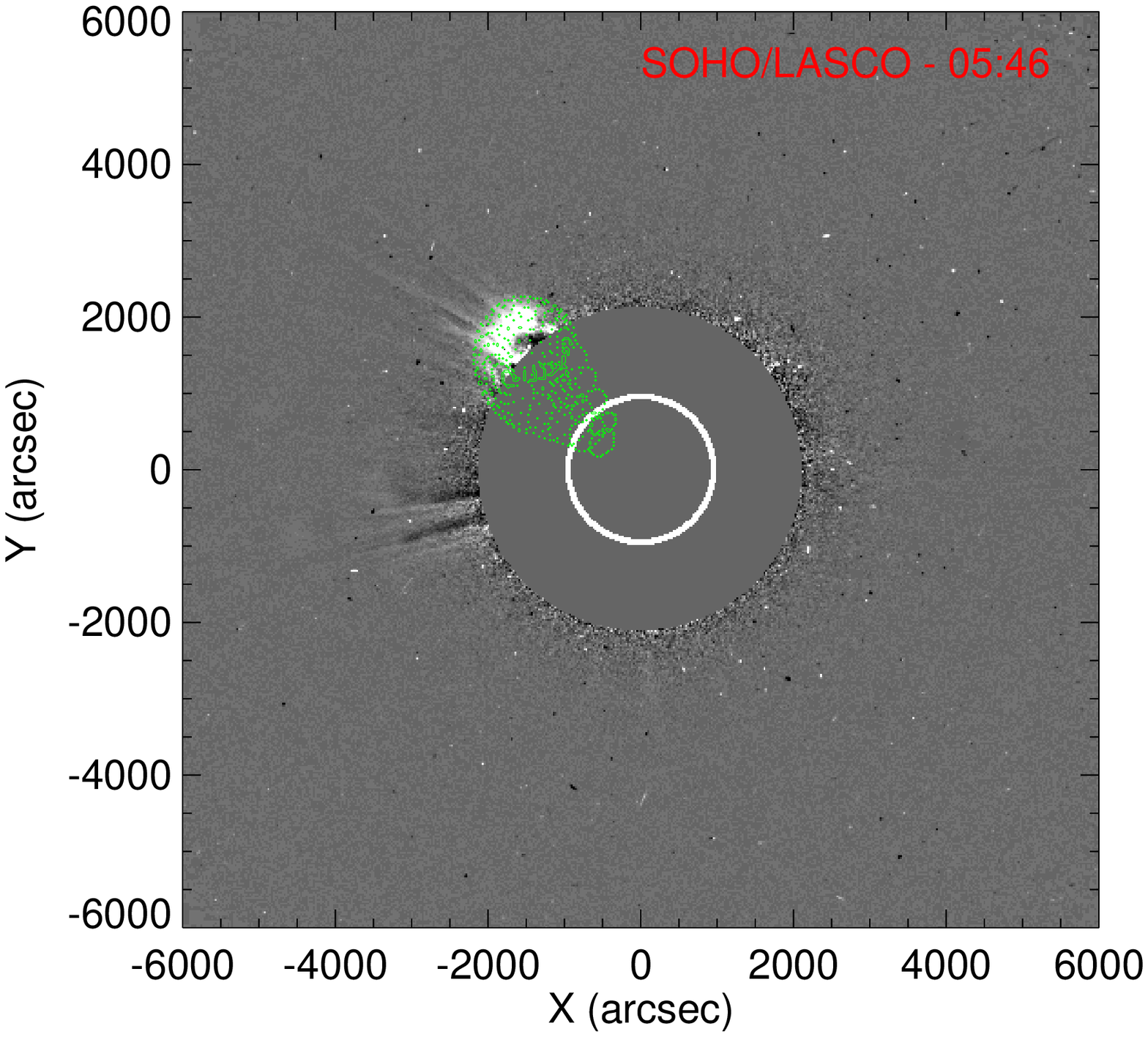}
    \includegraphics[width = 0.46\textwidth]{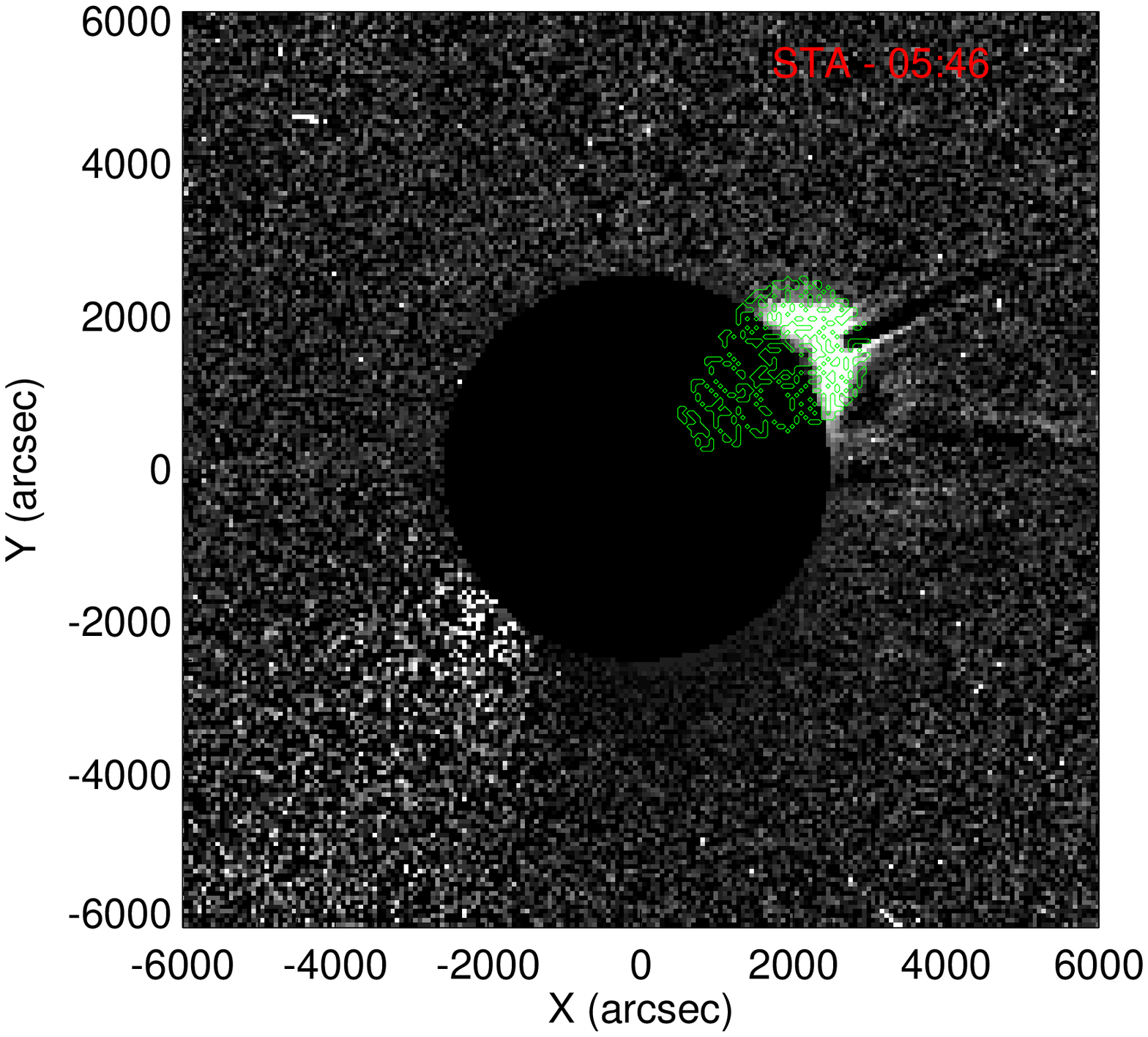}
    \caption{Top: The synoptic ecliptic-plane field plot for Carrington rotation 2217. The projected locations of STEREO-A (STA) and Earth are added as open circles and their magnetic footpoints mapped back to 2.5 solar radii based on the ballistic model are shown as filled circles. The outer edges of an EUV wave front at different times, i.e, at 05:00, 05:02, 05:07, 05:12, 05:17, 05:22 on May 6, are marked as colored dashed lines. Bottom: CME observation of SOHO/LASCO(left) and STA/Cor2(right) at 05:46 on May 6, 2019. The green mesh outlines the GCS reconstruction of the CME geometry. More description of this figure can be found in Sec.~\ref{sec:Remote-sensing}.}
    \label{fig:connection}
\end{figure}

At 05:00:20, an asymmetric EUV wave ($500$ km/s) started propagating toward the north-western hemisphere from the source as observed in the 193\AA~band of the Atmospheric Imaging Assembly (AIA) on board the Solar Dynamics Observatory (SDO).
The outer edges of the wave front at different times, i.e, at 05:00, 05:02, 05:07, 05:12, 05:17, 05:22 on May 6, are marked as colored dashed lines in the top panel of Fig.~\ref{fig:connection} which is explained in more detail below. After 05:22, the wave continued expanding across the solar surface as a rather faint structure which does not contain a clear wave front. 

About half an hour after the flare eruption, at 05:28, a CME first appeared in the field of view of the Large Angle and Spectrometric Coronagraph Experiment (LASCO) C2 of the \emph{Solar and Heliospheric Observatory} (SOHO) with a projected (plane of the sky) speed of 376 km/s and an angular width of 20$^\circ$, as reported in the CACTUS CME catalog\footnote{\url{http://sidc.oma.be/cactus/}}. 
Simultaneously, STA COR2 also captured the same structure. 
We applied the Graduated Cylindrical Shell (GCS) model \citep{GCS-model-Thernisien-2011ApJS} to the coronagraph observations from two directions to obtain the velocity and propagation direction of the CME. In the bottom panel of Fig.~\ref{fig:connection}, we give the CME observation and outline the fitted CME structure in green at 05:46 on May 6. The fitting results show that the linear speed of the CME front was about 326 km/s and the CME was deflected by about 10 degrees away from the location of the flare to the west, consistent with the direction of the EUV wave propagation.

In the top panel of Fig.~\ref{fig:connection}, we show a  magnetogram (in grey shades) on which the results of a potential field source-surface (PFSS) model for this Carrington rotation 2217 have been overlaid (source: Global Oscillation Network Group (GONG, \url{https://gong.nso.edu})). Open magnetic field lines (here with positive polarity shown in green and negative polarity shown in red) that connect the photosphere and the PFSS source surface for the solar wind are generally considered as the main channels for SEP propagation in the solar corona. The polarity inversion line is shown in blue. The above mentioned active region AR12470 is in the right part of the plot and marked by a white arrow. The radial projection points of STA and Earth on the solar surface on May 6 are plotted as open circles in black and blue, respectively. 
The magnetic footpoints based on the ballistic back mapping of the solar wind propagating at an average solar wind speed (about 360 km/s as observed in-situ at Earth) are marked as filled circles. 
We note that the magnetic footpoint of STA is only 7$^\circ$ away from the location of AR12470, which suggests STA is well connected to this active region. On the other hand, the longitudinal separation between the flare and Earth's footpoint is about 113$^\circ$ as displayed in Fig.~\ref{fig:connection}. Also, the wave doesn't appear to persist to the earth's footpoint.

\section{Summary and Discussion}

On 2019 May 6, an SEP event was observed by LND on the far-side surface of the Moon. While it had only a low intensity, this is nevertheless the first SEP event with enough counting statistics that is detected by LND. This event was also detected by \emph{SOHO}/EPHIN, which unfortunately only registered its decay phase, and also by \emph{Wind} and \emph{ACE}. The only possible solar source is a flare and its accompanying CME at AR 12470 located on the east hemisphere ($E50^\circ$). 
The type II radio burst indicates the existence of a shock in the lower corona and the CME speed was fitted as 326km/s by the GCS model.

The time profiles of electrons and protons clearly show velocity dispersion. According to the VDA analysis which assumes that all particles propagate along the same IMF line arriving at Earth and based on the combined data of LND, \emph{Wind} and \emph{ACE}, electrons were released about 22$\pm$3 minutes after the flare and type III radio burst and about 8 minutes after the high frequency type II radio burst, while protons were released at least 70 minutes later. 
The in-situ velocity dispersion and anisotropy of electrons suggest that a direct magnetic connection from the source to Earth was established for these electrons.
However, the separation between the flare location and the magnetic footpoint of Earth derived from the standard ballistic mapping is as large as 113$^\circ$.
This wide separation is remarkable.
In the classical scenario of particle transport during impulsive events \citep{Reames1999SSRv} accelerated particles stream along an open IMF connecting the source and the observer with high-energy particles arriving earlier than lower energy ones. The in-situ electron observation of this event at Earth (impulsive, velocity dispersion, beam-like distribution) suggests a good connection to the flare, 
while the remote sensing observation contradicts this explanation.

A possible explanation could be that particles are accelerated by the coronal shock which deviated westwards and reached the magnetic footpoint of Earth. For example, \citet{Rouillard2012ApJ-Longitudinal} have found an association between EUV waves which track the lateral shock expansion \citep{Veronig_2010} and particle release in wide-spread SEPs. As shown in the top panel of Fig.~\ref{fig:connection}, the EUV waves marked as colored dashed lines are propagating westward.
The GCS fitting of the CME also indicates a non-radial and westward-deflected propagation towards the location of the Earth footpoint. 
However, the deflection of the CME is only about 10$^\circ$ and the EUV wave became rather faint after 05:25 suggesting that the shock was unlikely to reach Earth's footpoint around the particle release time.
Therefore, it is difficult to conclude that a deflected shock should be responsible for the initial particle acceleration and release processes.

\citet{Klassen2018A&A-Strong} proposed another scenario in which electrons accelerated by a flare can reach a distant magnetic footpoint (90$^\circ$ in their example found with STA) through an irregular magnetic field at the solar source. 
The PFSS extrapolation of the magnetic field (top panel of Fig.~\ref{fig:connection}) before this eruption does not show any direct connection between the flare and the location of Earth's footpoint. The PFSS model, however, assumes a current-free field by definition, and is an idealized consideration of the solar corona, so the real magnetic configuration might be drastically different. Furthermore, solar eruptions often rearrange the solar magnetic fields through e.g., magnetic reconnection, and might have created a path for the particles to propagate over a large distance in heliolongitude. Unfortunately, the available observations do not show any evidence of a rearrangement of the solar magnetic field due to eruptions.

Moreover, the Parker spiral model is also an over-simplified, non-disturbed IMF condition and some researchers suggest the meandering and "random walk" of IMF would affect the particle propagation in the heliosphere \citep[e.g.,][]{Mazur-2000ApJ-Interplanetary,Laitinen2016A&A}.
In addition, the cross-field transport due to pitch-angle scattering and diffusion also causes particles to propagate in longitude and be observed on field lines with footpoints far away from the center of the solar eruption \citep[e.g.,][]{Wibberenz2006ApJ}. This means that one could observe particle events on poorly-connected field lines. However, the beam-like nature of the first electrons to arrive does not support such a cross-field transport model for this event.


Another observation about this event is the long delay between the electron and proton release times as derived from the VDA model which indicates that different acceleration processes and/or release locations might be responsible for protons and electrons. 
As protons are released much later, they may be more likely accelerated by the shock which, however, can not be confirmed to have reached Earth's magnetic footpoint. This explanation would also be difficult to reconcile with the common path lengths of electrons and protons, $L_e$ and $L_p$. 
In their study of the delay between the electron and proton onset times at STA, STB and L1 for wide-spread events, \citet{richardson2014} found that this time delay and the longitudinal separation between the flare location and backmapped spacecraft footpoint were correlated. In Tab.~\ref{tab:Time-line} we report a delay between the electron and protons release times which may contribute to the \citet{richardson2014} results. Such delays are not uncommon.



To summarize, we have presented various observations related to the first SEP event ever detected on the Lunar far-side surface. (1) The energy spectra of LND are consistent with observations from other spacecraft, though this is a weak event requiring large background subtractions. (2) The proton onset time of LND is also consistent with observations from other spacecraft, suggesting that the instrument response appears to be consistent with expectations. (3) The observations show clear velocity dispersion. (4) The SEP event was associated with a widely separated ($\sim 113^\circ$) flare at $E50^\circ$ relative to Earth.

\acknowledgments
The Lunar Lander Neutron and Dosimetry (LND) instrument is supported by the German Space Agency, DLR, and its Space Administration under grant 50 JR 1604 to the Christian-Albrechts-University (CAU) Kiel and supported by Beijing Municipal Science and Technology Commission, Grant No. Z181100002918003 and National Natural Science Foundation of China Grant: 41941001 to the National Space Science Center (NSSC). The scientific data are provided by China National Space Administration.
J. Guo and Y. Wang are supported by the Strategic Priority Program of the Chinese Academy of Sciences (Grant No. XDB41000000 and XDA15017300), and the CNSA pre-research Project on Civil Aerospace Technologies (Grant No. D020104).
N.D. was supported under grant 50OC1302 by the Federal Ministry of Economics and Technology on the basis of a decision by the German Bundestag.
We also would like to express our very great appreciation to Patrick K\"{u}hl, Lars Berger, Andreas Klassen and an anonymous referee for their valuable suggestions on this work.

\bibliography{main}{}



\appendix

\section{LND measurement and Poisson-CUSUM analysis}
\begin{figure}
    \centering
    \includegraphics[width=0.9\textwidth]{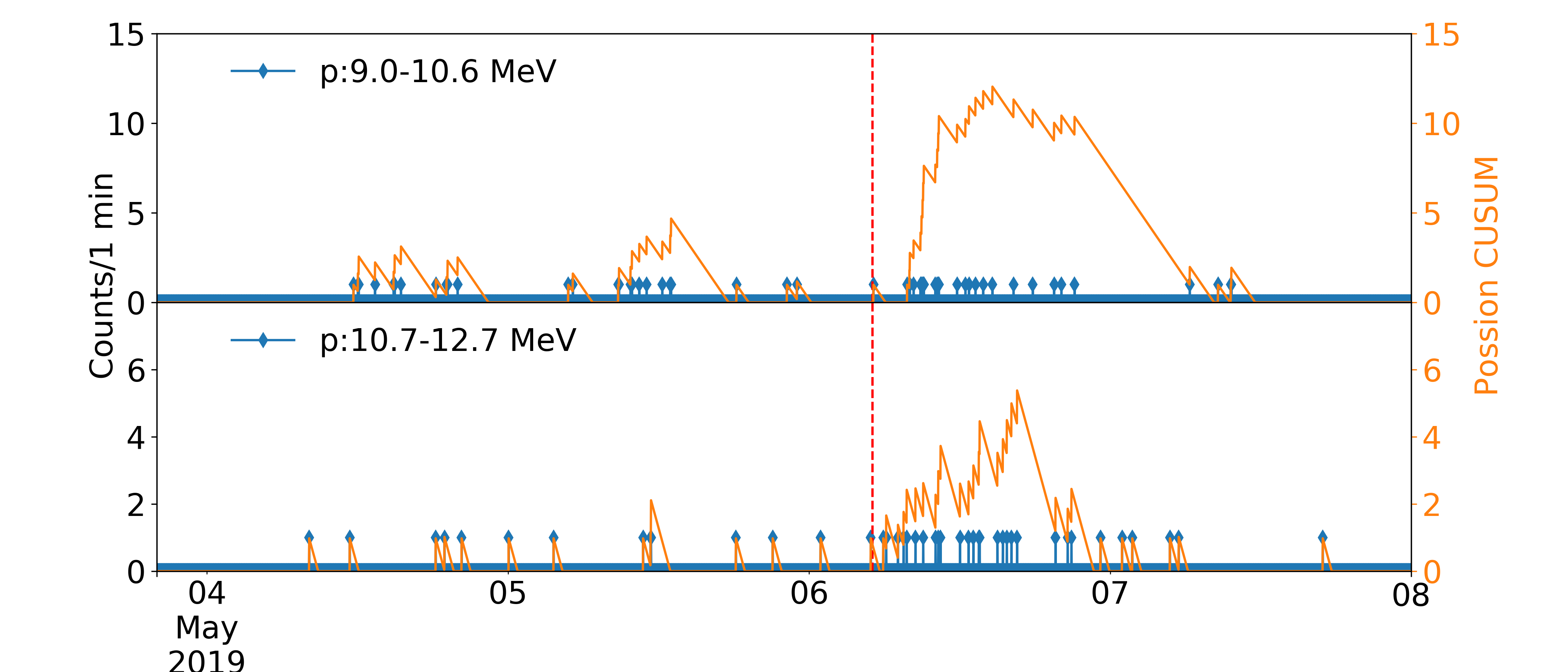}
    \caption{LND proton count rate in one minute time resolution(blue) and Poisson-CUSUM analysis(orange) based on one-minute data.}
    \label{fig:LND_counts_CUSUM}
\end{figure}

The charged particle telescope of LND measures protons in one-minute time resolution between 9.0--35 MeV and the explicit energy bins are given in Table. 6 of \cite{Wimmer-Schweingruber2020}. Here, we present the one-minute proton count rates at 9.0--10.6 MeV and 10.7--12.7 MeV during May 4 and May 8 in Fig.~\ref{fig:LND_counts_CUSUM}. The blue curves are the LND measurement in one-minute time resolution. Most of time, zero count is registered by LND, even during SEP event. One reason is the small intensity of the SEP event and the other is the small geometry factor of LND. Both of them cause the poor statistics.

In order to calculate the event onset time, the Poisson-CUSUM method is applied and the results are plotted in a orange line for each channel. Cumulative sum (CUSUM) control schemes are widely used in industrial applications because they are designed to give an indication of when there is a change in a process\citep{page1954}. In our case, the change is when the solar energetic particles rise above the background. The traditional CUSUM schemes are applied to a normally distributed quantity. When the variable has a Poisson distribution, a Poisson-CUSUM should be used. In the measurement of LND, the number of counts in a fixed interval, for example, one minute, obey a Poisson distribution. Hence, the Poisson-CUSUM will be used to determined the onset time of SEP event. By the definition of control schemes\citep{lucas1985-CUSUM}, the difference between the observed value $Y_{i}$ and a reference value $k$ are accumulated as a systematic change which is: 
\begin{equation}
    S_i  = max(0, Y_i - k + S_{i-1})
\end{equation}
The start values is $S_0$ = 0 in standard CUSUM.

The reference value $k$ is determined below:
\begin{equation}
    k = \frac{\mu_d - \mu_a}{ln(\mu_d) - ln(\mu_a)}
\end{equation}
where $\mu_a$ is the mean number of counts estimated for each channel during the pre-event background and $\mu_d$ is selected by a two-sigma-shift criterion\citep{Huttunen-Heikinmaa-2005A&A}:
\begin{equation}
    \mu_d = \mu_a + 2\sigma_a
\end{equation}
$\sigma_a$ is the standard deviation of the pre-event background. In order to have a non-zero background, the pre-event background is integrated from April 29 to May 2.

When the systematic change $S_i$ exceed the decision value h, then the onset of the SEP event is determined. Here we give a small decision value h = 1 according to the table in \cite{lucas1985-CUSUM}. In order to reduce the false alarms due to the small h, we apply the following criterion: once an $S_i$ exceeds the $h$, then the following 60 data points are checked as well. If all of them are larger than $h$, then the first signal is defined as the onset time of the SEP event.

We then apply the Poisson-CUSUM method for the one-minute data and the systematic changes are plotted as orange lines in Fig.~\ref{fig:LND_counts_CUSUM}. The red dashed vertical line marks the eruption time of flare at 04:56 on May 6. The increases of the orange lines during the SEP are very clear and the onset time can be easily determined. The onset time of the 9.0--10.6 MeV and 10.7--12.7 MeV channels are at 08:00 and 07:32 respectively.

We calculated the uncertainty of the Poisson-CUSUM method by determining the average time differences between the Poisson-CUSUM onset time and the times of the closest two neighboring non-zero data points.

The Poisson-CUSUM profiles, especially the one in 9.0-10.6MeV, indicate the existence of energetic particles registered by LND during May 4 and May 6, which are not explicitly shown in Fig.~\ref{fig:L1_measurement}(a). The corresponding lower energy protons are also detected by \emph{SOHO}/EPHIN and \emph{ACE}/EPAM. Those particles may have the same solar origin as that of the gradual SEP event detected by STA starting on May 4 and lasting until May 8, i.e., the ongoing event that we mentioned in Sec.~\ref{sec:In-situ measurementm} during May 6.

\end{document}